\documentclass[ ]{revtex4}
\usepackage{hyperref}
\usepackage{tikz}
\usepackage{amsmath}
\usepackage{amsfonts}
\usepackage{graphicx}
\begin{document}
\title{\bf \Large  Construction of a  Massive ABJM Theory Without Higgs Superfields }

 \author { Sudhaker Upadhyay }
 \email{ sudhakerupadhyay@gmail.com}
\affiliation { Centre for Theoretical Studies,\\
Indian Institute of Technology Kharagpur,  Kharagpur-721302, WB, India}

\begin{abstract}
A massive   Aharony-Bergman-Jafferis-Maldacena (ABJM) model in $\mathcal{N} =1$ superspace is analysed by considering
a Proca type mass term into the most general Faddeev-Popov action in a covariant gauge.
The presence of mass term breaks the original BRST and anti-BRST invariance of the model. Further, the symmetry of the massive ABJM model is restored by  extending the BRST 
and anti-BRST transformations. We show that the supergauge dependence of generating functional for connected diagrams occurs   in  presence of mass and ghost-anti-ghost condensates
in the theory. 
\end{abstract}

\maketitle
\section{Introduction}
According to the   AdS/CFT  correspondence,  certain gauge theories in   $d$-dimensions
correspond  to string/M-theory on backgrounds involving $d+1$ dimensional AdS spaces and vice versa. 
The M-theory was discovered due to the fact that eleven-dimensional supergravity arises as a low-energy limit of the ten-dimensional Type IIA superstring \cite{witt}. 
In fact, the detailed study of     AdS$_4$/CFT$_3$ correspondence on ABJM 
 theory \cite{abjm}  can be found  in Ref. \cite{mari}.
The  ABJM theory, a three-dimensional  (3D) ${\cal N}=6$ superconformal Chern-Simons theory  having
gauge group $U_k(N) \times U_{-k}(N)$  with bifundamental matter enjoying $SO(4)$ flavor symmetry,  is dual to  M-theory compactified on AdS$_4 \times S^7/{\mathbb{Z}}_k$, and describes the low energy dynamics of a stack
of M2 branes probing an orbifold singularity.
 In
particular, this correspondence is justified by the  planar free energy, which matches at strong coupling the classical IIA
supergravity action on AdS$_4$/CFT$_3$
and gives the correct 
$N^{3/2}$ scaling for the number of degrees of freedom of the M2 brane theory  \cite{68,32} .

In  M2-brane duality interpretation, the prediction that  the 3D superconformal field theory  should be the Chern-Simons gauge theory with maximal (${\cal N}=8$) supersymmetry  was first implemented by
 Bagger, Lambert and Gustavsson (BLG)  \cite{15,16}. The BLG theory uses 
algebraic structure known as Lie 3-algebras (and non-associative algebras).  Still, the  construction did not meet the desired dual to the M-theory
on AdS$_4 \times S^7$ as it  
works only for the gauge group $SO(4)$. 
Further,  Mukhi
 and   Papageorgakis examined the BLG theory for multiple M2-
branes and shown that when a scalar  field   in the 3-algebra develops a vacuum 
expectation value, the resulting Higgs mechanism has the novel effect  of topological   to dynamical   gauge  fields promotion \cite{mu,19}. This novel Higgs mechanism  is used to
   determine the leading
higher-derivative corrections to the maximally supersymmetric BLG  and ABJM 
theories. 
In each case, these superconformal theories are related, through the
novel Higgs mechanism, to the Yang-Mills theory on D2-branes.  
A massive Yang-Mills theory  via   the Higgs mechanism, in which local
gauge invariance is spontaneously broken by the
Higgs field and, thus, a gauge field acquires mass,  satisfies both renormalizability and physical unitarity \cite{18}. 
 A mass deformation of the Bagger-Lambert theory without
breaking any supersymmetry is studied in Ref. \cite{oo} where a mass-deformed theory is one example of
the 3D supersymmetric field theory with the so-called `non-central' term
whose superalgebra has been studied before \cite{m,n}. 
Seeking the importance of the massive superconformal Chern-Simons theory, we try to provide 
a  massive construction of the ABJM theory without introducing Higgs superfield.

A non-perturbative construction of massive Yang-Mills fields without introducing 
the Higgs field is studied recently   \cite{kon}, where   renormalizability and physical unitarity could not be established. 
A conventional argument for the violation of physical unitarity in the massive Yang-Mills
theory without the Higgs field is mentioned in \cite{kon1}. Specifically, it is shown that the
violation of physical unitarity follows from the incomplete
cancellation among unphysical modes: the scalar mode
with the Faddeev-Popov ghost and antighost. The obvious reason for the violation of physical unitarity is the non-nilpotency of BRST invariance due to mass term. 
 Subsequently, the   unitarity and renormalizability  for massive Yang-Mills fields in this 
 construction have been recovered by extending the BRST symmetry \cite{kon2}. 
 
 The BRST symmetry is one of the important tools to handle the gauge theory
consistently. The BRST symmetries for the superconformal Chern-Simons theories have been studied in recent past
with various motivations
 \cite{Mir, mir1, sud, sud1,sud001, sud2}. 
For instance,   the BRST symmetry, Slavnov-Taylor identities and Nielsen identities  are derived for the ABJM theories in N=3 harmonic superspace and the gauge dependence of one-particle irreducible amplitudes  is shown to be generated by a canonical flow with respect to the extended Slavnov-Taylor identity \cite{sud}. For a Delbourgo-Jarvis-Baulieu-Thierry-Mieg   type gauge, the spontaneous breaking of the BRST symmetry occurs in the BLG theory and  the responsible candidate for such spontaneous breaking is ghost-anti-ghost condensation \cite{sud1}. The generalized BRST symmetry,  by making the transformation parameter finite and field-dependent,
also known as finite field-dependent BRST transformation \cite{sud001}, is  
discussed for the superconformal Chern-Simons theories \cite{sud2}.

Though original superconformal Chern-Simons theories are maximally  supersymmetric,   we 
consider a particular (gauge) sector of those supersymmetric
gauge theories by using $\mathcal{N} =1$ superfields in
three dimensions for simplicity. In order to remove the redundancy
in gauge degrees of freedom, the Faddeev-Popov action for the $\mathcal{N} =1$  ABJM theory
 is constructed in the  most general covariant gauge.  The resulting Faddeev-Popov action 
respects absolutely anti-commuting BRST and anti-BRST transformation on  Curci-Ferrari
(CF) restricted surface. A Proca type mass term is added to the action, which breaks the
gauge invariance. Therefore, this breaks the BRST and anti-BRST symmetries as well. To restore the symmetry, we extend the BRST and anti-BRST symmetry
transformations in such a manner that these leave the massive ABJM theory invariant.
But the cost we pay is that such extended transformations are not nilpotent. The responsible candidate, for 
breaking the nilpotency, is mass parameter $M$.   Not only  nilpotency, 
also the   anticommutativity
is lost due to $M$.
Furthermore,  we check the exactness of the
gauge-fixed (originally BRST-exact) action under extended BRST and anti-BRST transformation  
and found  that   the presence of either  mass $M$, gauge   parameter $\alpha$ or
condensate $\mbox{Tr} [\bar c c- \tilde{\bar c}\tilde c]$ breaks the 
extended-BRST exactness. 
Following the Kugo-Ojima subsidiary condition, for extended BRST and anti-
BRST transformations,  the gauge dependence of the generating functional of
the connected Green functions   is analysed. We found that
the generating functional depends on the  gauge parameter
 only if  mass $M \neq 0$ and the condensate $\mbox{Tr}[\bar c c- \tilde{\bar c}\tilde c]\neq 
 0$.  Further, the construction of the massive gauge superfield connections without
 Higgs superfields is made. With the help of these massive  gauge superfields, the off-shell 
 and on-shell extended BRST invariant condensates are also computed.

 The presentation of this paper is as following. In section II,
 we recapitulate the construction of the $\mathcal{N} =1$  ABJM theory  in three dimensional  
superspace. The Faddeev-Popov treatment for this theory in most general covariant
gauges is presented in section III. The massive theory without Higgs superfields and the
extended BRST and anti-BRST symmetry are  
discussed in section IV. The gauge dependence of generating functional is  
elucidated. The construction of massive supergauge connections is presented in
section V. The results with future motivations are reported in the last section.
 
\section{  ABJM Theory: preliminaries }
We discuss the preliminaries of the  $\mathcal{N} =1$  ABJM theory  in 3D
superspace 
$(x^\mu, \theta_a)$, where $\theta$ is a
Grassmann spinor. We   adopt the following notation for 3D superspace \cite{keto}:
\begin{eqnarray}
\partial_\mu =\frac{\partial}{\partial x^\mu},\ \ \ \partial_a =\frac{\partial}{\partial \theta^a}.
\end{eqnarray}
The $\mathcal{N} =1$  supersymmetry generators are conveniently represented
in 3D superspace by
 $Q_a = \partial_a -i(\gamma^\mu \theta)_a \partial_\mu$, and follow
 $\{Q_a , Q_b \}= 2i\varepsilon_{bc}(\gamma^\mu)_{a}^{\ c}\partial_\mu$.
The generators   $Q_a$ also satisfy $\{D_a, Q_b\} =0$, where    
$D_a = -i\partial_a + (\gamma^\mu \theta)_a \partial_\mu$ denote supersymmetrically invariant  derivatives. The  supercovariant
derivatives can be defined by covariantizing
$D_a$ as follows:
\begin{eqnarray}
\nabla_a =D_a +i\Gamma_a,
\end{eqnarray}
where the supergauge connection $\Gamma_a$ belongs to the adjoint
representation of one of the gauge groups  $U(N)  \times U (N)$. These supercovariant
derivatives lead to the following off-shell superfield constraints:
$\{\nabla_a,\nabla_b\}=-2i\nabla_{ab}$.
The  non-Abelian supercovariant
superfield strength, $\Omega$, related to  supercovariant
derivatives is computed by
\begin{eqnarray}
[\nabla_a, \nabla_{bc}] =-\varepsilon_{a(b}\Omega_{c)},
\end{eqnarray}
with  
\begin{eqnarray}
 \Omega_a =  \frac{1}{2} D^b D_a \Gamma_b - \frac{i}{2} 
 [\Gamma^b , D_b \Gamma_a]    -
 \frac{1}{6} [ \Gamma^b ,
\{ \Gamma_b , \Gamma_a\}    ].  
\end{eqnarray}
In component form, spinor superfield  can be expressed  as  \cite{keto}
\begin{eqnarray}
 \Gamma_a = \xi_a + \frac{1}{2} \theta_a G +  (\gamma^\mu\theta)_a A_\mu + i\theta^2 \left[\lambda_a -
 \frac{1}{2}(\gamma^\mu \partial_\mu \xi)_a\right].\label{gam}
  \end{eqnarray}
 The  action  for the  $\mathcal{N} =1$ ABJM theory  with the gauge group $U(N)_{k}  \times U(N)_{-k}$ is given by
\begin{equation}
{ S} =  S_{\mbox{matter}} + S_{\mbox{CS}},\label{lag}
\end{equation} 
where   $S_{\mbox{CS}} $   is the super-Chern-Simons action  in an adjoint representation  with the
following explicit form:
 \begin{eqnarray}
 S_{\mbox{CS}} &=& \frac{k}{16\pi} \int d^3x d^2 \,  \theta \, \, 
 \mbox{Tr} \left[i  \Gamma^a         \Omega_a +\frac{1}{6}\{\Gamma^a,\Gamma^b\}D_b\Gamma_a +\frac{i}{12}\{\Gamma^a,\Gamma^b\}\{\Gamma_a,\Gamma_b\} \right.\nonumber\\
 &-& \left. i\tilde{\Gamma}^a         \tilde{\Omega}_a -\frac{1}{6}\{\tilde\Gamma^a,\tilde\Gamma^b\}D_b\tilde\Gamma_a -\frac{i}{12}\{\tilde\Gamma^a,\tilde\Gamma^b\}\{\tilde\Gamma_a,\tilde\Gamma_b\}
\right].
\end{eqnarray}
Here $k$ is an integer and known as label. Corresponding to gauge groups $U(N)_{k}$ and 
$U(N)_{-k}$, we   introduce  different   supergauge connections
$\Gamma^a$ and $\tilde{\Gamma}^a$, respectively. 
Analogously to (\ref{gam}), the  non-Abelian supercovariant
superfield strength involving $\tilde{\Gamma}^a$ is given by
 \begin{eqnarray}
\tilde \Omega_a & = &\frac{1}{2} D^b D_a \tilde \Gamma_b 
- \frac{i}{2}  [\tilde \Gamma^b , D_b \tilde\Gamma_a]    -
 \frac{1}{6} [ \tilde \Gamma^b ,
\{ \tilde \Gamma_b ,  \tilde \Gamma_a\} ].
\end{eqnarray}  
 The matter   action is given by \cite{keto},
\begin{eqnarray}
S_{\mbox{matter}} &=& \frac{1}{4} \int d^3x d^2 \,  \theta \, \,  
\mbox{Tr}  \left[  \nabla^a_{}         X^{I \dagger}               
\nabla_{a 
}               X_I      \right],
\end{eqnarray}
where the matrix-valued complex scalar matter superfield $X_I$  is in the bi-fundamental 
representation
of the gauge group. 
 
Under the supergauge
transformations, the connections and matter superfields change  as follows,
\begin{eqnarray}
 \delta \Gamma_a =  \nabla_a {     } \Lambda =D_a\Lambda +i[\Gamma_a,\Lambda], 
&&   \delta \tilde\Gamma_a = \tilde\nabla_a {     }
 \tilde\Lambda  =D_a\tilde\Lambda +i[\tilde\Gamma_a,\tilde\Lambda], \nonumber \\ 
\delta X^{I } = i(\Lambda{     } X^{I }  - X^{I }{     }\tilde \Lambda ),  
&&  \delta  X^{I \dagger  } 
= i(   \tilde \Lambda {     } X^{I\dagger  }-X^{I\dagger  }{     } \Lambda), 
\end{eqnarray}
where $\Lambda$ and $\tilde\Lambda$ are the gauge Lie-algebra-valued parameters.

   Under these  supergauge transformations, the ${\cal N}=1$ ABJM action  (\ref{lag}) 
   is invariant. It is well-known 
   that the redundancies in supergauge degrees of freedom are associated 
   with the gauge symmetry. So, in the next section, we would try to remove them by quantizing the ABJM model using Faddeev-Popov    method.
\section{ The ABJM Theory: BRST symmetry }
 In order to remove unphysical supergauge degrees of freedom, we break the local supergauge
  invariance  with the supergauge conditions:
  $
  D^a \Gamma_a  =0$ and $ D^a \tilde{\Gamma}_a =0
$. Now, the effect  of such conditions is incorporated in the theory  by adding 
the following (most general) Faddeev-Popov term to the supergauge invariant 
action:
\begin{eqnarray}
S_{\mbox{gf+gh}} &= &\int d^3x d^2 \,  \theta \, \,  \mbox{Tr}   \left[ B  (D^a \Gamma_a) + \frac{\alpha }{2}BB -i\frac{\alpha }{2}  B[\bar c, c] +i\bar c D^a\nabla_a c -\frac{\alpha }{4}
[\bar c, c][\bar c, c]  \right.\nonumber\\
& -&\left.
  \tilde{B}    (D^a \tilde{\Gamma}_a) - \frac{\alpha }{2}\tilde{B}    \tilde B 
  +i\frac{\alpha }{2} \tilde{B}  [\tilde{\bar c}, \tilde c] -i\tilde{\bar c} D^a \tilde \nabla_a \tilde c +\frac{\alpha }{4}
[\tilde{\bar c}, \tilde c][\tilde{\bar c}, \tilde c]   
\right],\label{gf}
\end{eqnarray}
where $\alpha$ is a supergauge parameter. Here  Nakanishi-Lautrup type superfields $\bar B$ and $B$,  $\tilde{\bar B}$ and $\tilde B$ are related with the following 
 CF-type restrictions :
\begin{eqnarray}
\bar{B}&=&-B +i [\bar c, c],\nonumber\\
\tilde{\bar{B}}&=&-\tilde B +i [\tilde{\bar c}, \tilde c].\label{cf}
\end{eqnarray}
This can be  further    recast  as
\begin{eqnarray}
S_{\mbox{gf+gh}}  
 &= &\int d^3x  d^2 \,  \theta \, \,  \mbox{Tr}   \left[ B  (D^a \Gamma_a) +i\bar c D^a\nabla_a c  +
 \frac{\alpha }{4} \left(\bar B\bar B+ BB -\tilde{\bar B}\tilde{\bar B}- \tilde B
 \tilde B\right)\right.\nonumber\\
& -&\left.
  \tilde{B}    (D^a \tilde{\Gamma}_a)   -i\tilde{\bar c} D^a \tilde \nabla_a \tilde c  
\right].
\end{eqnarray}
Now, the resulting effective (Faddeev-Popov) action, $S+S_{\mbox{gf+gh}}$, is invariant under 
the following sets of BRST ($s_{b}$) and anti-BRST ($s_{ab}$) transformations :
\begin{eqnarray}
&&s_b \,\Gamma_{a} = \nabla_a    c,\ \   \   s_b \, \tilde\Gamma_{a} = \tilde\nabla_a    
 \tilde c, \  \ \
 s_b  \,c = -  \frac{1}{2}  {[c,c]}_ { },\ \  \ s_b  \,\tilde{ {c}} = -  \frac{1}{2}  [\tilde{ {c}} ,  \tilde c]_{ }, \nonumber \\
 &&s_b  \,\bar{c} =   iB,\    \ \  s_b  \,\tilde {\bar c} =  i\tilde B, \  \  
 s_b  \,B =0,\   \  s_b  \, \tilde B= 0, \  \ \ 
C \, X^{I } = i    c   X^{I } -  iX^{I }  \tilde c,\nonumber \\ 
 && 
  s_b  \, X^{I \dagger }
 = i   \tilde c    X^{I \dagger } - i  X^{I \dagger }  c,\label{brs}
\end{eqnarray}  
   \begin{eqnarray}
&&s_{ab} \,\Gamma_{a} = \nabla_a   \bar c,\  \ \ \   s_{ab}\, \tilde\Gamma_{a} = \tilde\nabla_a    
 \tilde{\bar c}, \ \ \
  s_{ab}  \,\bar c = -  \frac{1}{2}  {[\bar c, \bar c]}_ { },\ \  \ s_{ab}  \,\tilde{\bar {c}} = -  \frac{1}{2}  [\tilde{\bar {c}} ,  \tilde {\bar c}]_{ }, \nonumber \\
 &&s_{ab}  \, {c} =   i\bar B,\ \  \  s_{ab}  \,\tilde { c} =  i\tilde{\bar B}, \  \  
 s_{ab}  \,\bar B =0,\ \ \  \  s_{ab}  \, \tilde {\bar B}= 0, \  \  
 s_{ab}  \, X^{I } = i    \bar c   X^{I } -  iX^{I }  \tilde{\bar c},\nonumber \\ 
 && 
  s_{ab} \, X^{I \dagger }
 = i   \tilde{\bar c}    X^{I \dagger } - i  X^{I \dagger } \bar c.\label{abrs}
\end{eqnarray}  
It is easy to verify that the above transformations are nilpotent, i.e, $(s_b)^2= (s_{ab})^2=0$,
as well as absolutely anti-commuting in nature with CF restrictions (\ref{cf}), i.e., $\{s_a, s_{ab}\}=0$. 
\section{ The Massive ABJM Theory: Extended BRST symmetry }
To construct an effective  massive ABJM theory without 
the Higgs superfields,  we introduce following Proca-type mass term :
 \begin{eqnarray}
S_{\mbox{mass}} &= &M^2\int d^3x d^2 \,  \theta \, \,  \mbox{Tr}   \left[  \frac{1}{2} \Gamma_a\Gamma^a
+\alpha   (i\bar c c) - \frac{1}{2}  \tilde\Gamma_a\tilde \Gamma^a
-\alpha   (i\tilde{\bar c} \tilde c)   
\right].
\end{eqnarray}
With this mass term, the action results in
\begin{eqnarray}
S_{\mbox{eff}}=S+S_{\mbox{gf+gh}}+S_{\mbox{mass}},\label{me}
\end{eqnarray}
which, in turn, is not invariant under the BRST transformation (\ref{brs}) and anti-BRST transformation  (\ref{abrs}), as
\begin{eqnarray}
s_b(S_{{\mbox{mass}}})  \neq 0,\ \ \ s_{ab}(S_{{\mbox{mass}}})  \neq 0. 
\end{eqnarray}
 To restore the invariance of the action $S_{\mbox{eff}}$,  we extend the BRST transformation  (\ref{brs}) and anti-BRST transformation  (\ref{abrs}). 
 The extended BRST transformation (${s}^{m}_b $) and anti-BRST transformation (${s}^{m}_{ab}$), respectively,  are given by
  \begin{eqnarray}
&&{s}^{m}_b \,\Gamma_{a} = \nabla_a    c,\  \   s^{m}_b \, \tilde\Gamma_{a} = \tilde\nabla_a    
 \tilde c, \  \ 
 s^{m}_b  \,c = -  \frac{1}{2}  {[c,c]}_ { },\  \ s^{m}_b  \,\tilde{ {c}} = -  \frac{1}{2}  [\tilde{ {c}},  \tilde c]_{ }, \nonumber \\
 &&s^{m}_b  \,\bar{c} =   iB,  \ \  s^{m}_b  \,\tilde {\bar c} =  i\tilde B, \ \ 
 s^{m}_b  \,B =M^2c,\   \  s^{m}_b  \, \tilde B= M^2\tilde c, \nonumber \\ 
 &&s^{m}_b  \, X^{I } = i    c   X^{I } -  iX^{I }  \tilde c,\ \ 
  s^{m}_b  \, X^{I \dagger }
 = i   \tilde c    X^{I \dagger } - i  X^{I \dagger }  c,\label{brs1}
\end{eqnarray}  
  and
  \begin{eqnarray}
&&s^{m}_{ab} \,\Gamma_{a} = \nabla_a   \bar c,\  \   s^{m}_{ab}\, \tilde\Gamma_{a} = \tilde\nabla_a    
 \tilde{\bar c}, \ \ 
 s^{m}_{ab}  \,\bar c = -  \frac{1}{2}  {[\bar c, \bar c]}_ { },\    \ s^{m}_{ab}  \,\tilde{\bar {c}} = -  \frac{1}{2}  [\tilde{\bar {c}} ,  \tilde {\bar c}]_{ }, \nonumber \\
 &&s^{m}_{ab}  \, {c} =   i\bar B,\  \  s^{m}_{ab}  \,\tilde { c} =  i\tilde{\bar B},  \ \ s^{m}_{ab}  \,\bar B =-M^2 \bar c,\ \  s^{m}_{ab}  \, \tilde {\bar B}= -M^2 \tilde{\bar c}, \nonumber \\ 
 &&s^{m}_{ab}  \, X^{I } = i    \bar c   X^{I } -  iX^{I }  \tilde{\bar c},\ \ 
  s^{m}_{ab} \, X^{I \dagger }
 = i   \tilde{\bar c}    X^{I \dagger } - i  X^{I \dagger } \bar c.\label{abrs1}
\end{eqnarray} 
It is observed that  the extended BRST and anti-BRST transformations leave 
the massive effective action (\ref{me}) invariant, however the
mass term is invariant under these as 
\begin{eqnarray}
s^m_b(S_{{\mbox{mass}}}) \neq 0,\ \ \ s^m_{ab}(S_{{\mbox{mass}}}) \neq 0.
\end{eqnarray}
The Nakanishi-Lautrup superfields transform under 
extended BRST and anti-BRST transformations  as
  \begin{eqnarray}
  s^m_b\bar B&=&[\bar B, c] -M^2 c,\ \
  s^m_b\tilde{\bar B}= [\tilde{\bar{B}}, \tilde{c}] -M^2\tilde{c},\nonumber\\
  s^m_{ab} B&=& [B,\bar{c}] +M^2 \bar c,\ \
  s^m_{ab}\tilde{B}= [\tilde{B}, \tilde{\bar{c}}]+M^2\tilde{\bar{c}}.
  \end{eqnarray}
  Here, we have utilized the CF conditions (\ref{cf}).

To see the nilpotency of the extended BRST and anti-BRST transformations,
we apply these transformations twice on each superfields   and 
find the following non-vanishing  superfields:
  \begin{eqnarray}
  (s_b^{m})^2 \bar c= iM^2 c,\ \
(s_b^{m})^2 \tilde{\bar c}= iM^2 \tilde c,\ \
(s_b^{m})^2 B= -M^2 [c, c],\ \
  (s_b^{m})^2 \tilde B= -M^2 [\tilde c, \tilde c],
  \end{eqnarray}
  and 
   \begin{eqnarray}
  (s_{ab}^{m})^2 c= iM^2 \bar c,\ \
  (s_{ab}^{m})^2 \tilde{c}= iM^2 \tilde {\bar c}, \
 (s_{ab}^{m})^2 \bar B= -M^2 [\bar c, \bar c],\ \
   (s_{ab}^{m})^2 \tilde {\bar B}= -M^2 [\tilde {\bar c}, \tilde {\bar c}].
  \end{eqnarray}
It eventually confirms that the extended BRST and anti-BRST transformations are not nilpotent. However, in   $M\rightarrow0$ limit, the nilpotency of extended transformations
is evident, which is obvious as these transformations in   massless limit 
   correspond to the usual BRST and anti-BRST transformations. 
Due to  mass parameter, these extended BRST and anti-BRST transformations are
not absolutely anti-commuting i.e.  
 \begin{eqnarray}
  \{s_b^m, s_{ab}^m\}\neq 0,
  \end{eqnarray}
even on account of CF type restrictions.

The gauge-fixed action together with ghost term (\ref{gf}) is BRST-exact and can be expressed as
   \begin{eqnarray}
S_{\mbox{gf+gh}}&=& -s_b\int d^3x d^2\theta \mbox{Tr}\left[i\bar c\left(D^a\Gamma_a +\frac{\alpha }{2}B -
i\frac{\alpha }{4}[\bar c, c]\right) - i\tilde{\bar c}\left(D^a\tilde\Gamma_a +\frac{\alpha }{2}\tilde 
B -i\frac{\alpha }{4}[\tilde{\bar c}, \tilde c]\right)\right].
 \end{eqnarray}
Moreover, this  action (\ref{gf}) is not exact under the extended BRST and anti-BRST transformations, which is evident from the following:
  \begin{eqnarray}
S_{\mbox{gf+gh}}&=& -s^m_b\int d^3x d^2\theta \mbox{Tr}\left[i\bar c\left(D^a\Gamma_a +\frac{\alpha }{2}B -
i\frac{\alpha }{4}[\bar c, c]\right) - i\tilde{\bar c}\left(D^a\tilde\Gamma_a +\frac{\alpha }{2}\tilde 
B -i\frac{\alpha }{4}[\tilde{\bar c}, \tilde c]\right)\right]\nonumber\\
&-& i\frac{\alpha }{2}\int  d^3x d^2\theta \mbox{Tr} \left[ \bar c s_b^m B -\tilde{\bar c}s_b^m \tilde B \right],\nonumber\\
&=& -s^m_b\int d^3x d^2\theta \mbox{Tr}\left[i\bar c\left(D^a\Gamma_a +\frac{\alpha }{2}B -
i\frac{\alpha }{4}[\bar c, c]\right) - i\tilde{\bar c}\left(D^a\tilde\Gamma_a +\frac{\alpha }{2}\tilde 
B -i\frac{\alpha }{4}[\tilde{\bar c}, \tilde c]\right)\right] \nonumber\\
&-& iM^2\frac{\alpha }{2}\int d^3x d^2\theta \mbox{Tr} \left[ \bar c c -\tilde{\bar c} \tilde c\right],\nonumber\\
&=& is^m_bs^m_{ab}\int  d^3x d^2\theta \mbox{Tr}\left[\frac{1}{2}\Gamma^a \Gamma_a  +\frac{\alpha }{2} \bar c  
c -\frac{1}{2}\tilde\Gamma^a \tilde\Gamma_a -\frac{\alpha }{2} \tilde{\bar c}\tilde  c  
 \right]-iM^2\frac{\alpha }{2}\int  d^3x d^2\theta \mbox{Tr} \left[ \bar c c -\tilde{\bar c} 
 \tilde c\right].
\end{eqnarray}
The responsible features for this  non-exactness 
are non-vanishing mass $M$, parameter $\alpha $ and condensate 
$\mbox{Tr} \left[\bar c c -\tilde{\bar c} \tilde c\right]$.
 
The invariance of the  massive ABJM action under the
extended BRST transformation is justified from the
following computations: 
\begin{eqnarray}
s_b^m S_{\mbox{gf+gh}}&=&\int d^3x d^2\theta\ \mbox{Tr}\left[ (s_b^m B)D^a\Gamma_a +\alpha  
(s_b^mB)B -i\frac{\alpha }{2}(s_b^m B) [\bar c, c]\right.\nonumber\\
&-& \left. (s_b^m \tilde B)D^a\tilde\Gamma_a -\alpha  (s_b^m\tilde B)\tilde B +i\frac{\alpha 
}{2}(s_b^m \tilde B) [
\tilde{\bar c}, \tilde c] \right],\nonumber\\ 
&=&-M^2s_b^m \int  d^3x  d^2\theta\ \mbox{Tr}\left[\frac{1}{2}\Gamma^a\Gamma_a +i\alpha  \bar 
c c -\frac{1}{2}\tilde\Gamma^a\tilde\Gamma_a -i\alpha  \tilde{\bar c}\tilde c  \right],
\nonumber\\
&=& -s_b^m S_{\mbox{mass}}.
\end{eqnarray} 
Consequently, 
\begin{eqnarray}
s_b^m (S_{\mbox{gf+gh}}+S_{\mbox{mass}}) =0.
\end{eqnarray}
The classical action remains invariant under extended BRST transformation as it does not depend on Nakanishi-Lautrup superfields.

To study the supergauge dependence, we first define  the vacuum functional  $Z[J]$
with a source $J$ as follows,
\begin{eqnarray}
Z[J]=\int {\cal D}\Phi \exp\left[{iS_{\mbox{eff}}+\int d^3xd^2\theta\ \mbox{Tr} J {\cal A}}\right]=
 e^{iW[J]},
\end{eqnarray}
where ${\cal D}\Phi$ refers to a generic functional measure, ${\cal A}$ is an operator, and
 $W[J]$ represents the generating functional of the connected
Green functions of the massive ABJM theory.   The expectation value for the operator  ${\cal A}$ 
is given by
\begin{eqnarray}
\langle {\cal A} \rangle =\frac{\delta}{\delta J} W[J]|_{J=0}.
\end{eqnarray}
Now, to see the dependence of $W[J]$ on $\alpha $, we differentiate $W[J]$ 
 with respect to $\alpha$ and obtain,
\begin{eqnarray}
\frac{\partial W[J]}{\partial\alpha }=\frac{M^2}{2}\int d^3x d^2\theta\ \mbox{Tr}\langle   i  \bar c c  -i  \tilde{\bar c}\tilde c  \rangle \neq 0.\label{30}
\end{eqnarray}
Here,   the  Kugo-Ojima subsidiary condition corresponding to  extended 
BRST  transformations is adopted, i.e., the conserved  charges
for such transformations will annihilate the physical states of the total Hilbert states.  
From   expression (\ref{30}), it is evident that  $W[J]$ depends on the parameter $\alpha $
for  mass $M \neq 0$ together with condensate $ \mbox{Tr}\langle   i  \bar c c  -i  \tilde{\bar c}\tilde c  \rangle \neq 0$. {  In the massive limit,  $\alpha$
becomes a physical parameter defining   a mass  of 
anticommuting fields $c,\tilde{c},\bar c$ and $\tilde{\bar c}$ in the form  $\alpha M^2$ which can be seen from the equations of motion. To determine whether such ghost-anti-ghost condensation  
occurs or not, , it is important
to evaluate the effective potential for the composite operator (see, e.g., \cite{sud1} for details). 
This result should be compared with the massless case, in
which, on contrary, $W[J]$ does not depend on a gauge-fixing parameter $\alpha$. 
This implies that, for $M=0$,  any choice
of gauge-fixing parameter $\alpha$ gives the same  generating functional $W[J]$.  
It is no wonder that there is no dependence on this parameter because in this
limit the  effective action reduces to the Faddeev-Popov action and the nilpotency of the BRST transformations is restored.}
\section{Massive ABJM Superfields without Higgs Superfields}
In this section, we construct the massive supergauge superfields $\mathcal{W}_a$ and $\tilde{\mathcal{W}}_a$. The requirement for the physical massive vector superfields are: (i) these superfields must belong to the physical field creating a physical state with positive norm, (ii) these superfields should 
 have the correct degrees of freedom as a massive supergauge particle, and (iii) these superfields must obey the same transformation
rule as that of the original supergauge superfields.

Keeping these points in mind,   $\mathcal{W}_a$ and $\tilde{\mathcal{W}}_a$ are constructed 
by a nonlinear  local transformations  as follows,
\begin{eqnarray}
\mathcal{W}_a&=&\Gamma_a -\frac{1}{M^2}D_a B -\frac{1}{M^2}[\Gamma_a, B]
 +  \frac{i}{M^2}[D_a c, \bar c] + \frac{i}{M^2}[[\Gamma_a, c], \bar c],\nonumber\\
 &=&\Gamma_a +\frac{1}{M^2} is_b^ms_{ab}^m \Gamma_a,\label{cc}\\
 \tilde{\mathcal{W}}_a&=&\tilde\Gamma_a -\frac{1}{M^2}D_a \tilde B -\frac{1}{M^2}[\tilde
 \Gamma_a, \tilde B]
 +  \frac{i}{M^2}[D_a \tilde c, \tilde{\bar c}] + \frac{i}{M^2}[[\tilde\Gamma_a, 
 \tilde c], \tilde{\bar c}],\nonumber\\
 &=&\tilde\Gamma_a +\frac{1}{M^2} is_b^ms_{ab}^m \tilde\Gamma_a.\label{dd}
\end{eqnarray}
{These superfields fulfill the requirements discussed above
as (i) they have the modified BRST invariance (off-mass-shell), i.e.,
${s}^{m}_b \,\mathcal{W}_a = 0, \  s^{m}_b \, \tilde{\mathcal{W}}_a = 0$,
 (ii) $\mathcal{W}_a$ and $\tilde{\mathcal{W}}^a $ are divergenceless (on-mass-shell),
 i.e. $D_a \,\mathcal{W}^a = 0, \ D_a  \tilde{\mathcal{W}}^a = 0$.}

With the help of  expressions (\ref{cc}) and (\ref{dd}), we 
construct the Proca type mass terms $\frac{1}{2}M^2\mathcal{W}_a\mathcal{W}^a$  and 
 $\frac{1}{2}M^2 \tilde{\mathcal{W}}_a \tilde{\mathcal{W}}^a$,
which are invariant  
under the extended BRST transformations. These will be useful for the
 regularization scheme to avoid  divergences in the ABJM theory.  
Here, the   (off-shell) extended BRST invariant condensates  are  
\begin{eqnarray}
\langle \mathcal{W}_a\mathcal{W}^a\rangle,\ \ \ \langle \tilde{\mathcal{W}}_a\tilde{\mathcal{W}}^a\rangle,
\end{eqnarray}
and the   on-shell BRST invariant   condensates are
\begin{eqnarray}
\langle \frac{1}{2}\Gamma_a\Gamma^a +\alpha  c\bar c\rangle,\ \ \ \langle  \frac{1}{2}
\tilde \Gamma_a\tilde\Gamma^a +\alpha  \tilde c\tilde{\bar c}\rangle.
\end{eqnarray}
Therefore, the  massive effective ABJM action composed of
 massive vector fields $\mathcal{W}_a$ 
and $\tilde{\mathcal{W}}^a$ is invariant under the
following extended BRST and anti-BRST symmetry transformations, respectively:
  \begin{eqnarray}
&&{s}^{m}_b \,\mathcal{W}_a = 0,\ \ \ \  s^{m}_b \, \tilde{\mathcal{W}}_a = 0, \ \ \
  s^{m}_b  \,c = -  \frac{1}{2}  {[c,c]}_ { },   \ \ \ s^{m}_b  \,\tilde{ {c}} = -  \frac{1}{2}  [\tilde{ {c}},  \tilde c]_{ }, \nonumber \\
 &&s^{m}_b  \,\bar{c} =   iB, \ \ \  s^{m}_b  \,\tilde {\bar c} =  i\tilde B, \nonumber \ \ \ 
 s^{m}_b  \,B =M^2c,\   \ \  s^{m}_b  \, \tilde B= M^2\tilde c, \nonumber \\ 
 &&s^{m}_b  \, X^{I } = i    c   X^{I } -  iX^{I }  \tilde c,\ \  \
   s^{m}_b  \, X^{I \dagger }
 = i   \tilde c    X^{I \dagger } - i  X^{I \dagger }  c, 
\end{eqnarray} 
and 
   \begin{eqnarray}
&&s^{m}_{ab} \, {\mathcal{W}}^a = 0,\ \ \  s^{m}_{ab}\, \tilde{\mathcal{W}}^a = 0, \ \ \
 s^{m}_{ab}  \,\bar c = -  \frac{1}{2}  {[\bar c, \bar c]}_ { },\ \ \ \ \ \ \ \  \ \ s^{m}_{ab}  \,\tilde{\bar {c}} = -  \frac{1}{2}  [\tilde{\bar {c}} ,  \tilde {\bar c}]_{ }, \nonumber \\
 &&s^{m}_{ab}  \, {c} =   i\bar B,  \ \ \  s^{m}_{ab}  \,\tilde { c} =  i\tilde{\bar B}, \ \ \ 
 s^{m}_{ab}  \,\bar B =-M^2 \bar c,\  \ \  s^{m}_{ab}  \, \tilde {\bar B}= -M^2 \tilde{\bar c}, \nonumber \\ 
 &&s^{m}_{ab}  \, X^{I } = i    \bar c   X^{I } -  iX^{I }  \tilde{\bar c},\ \ \ 
  s^{m}_{ab} \, X^{I \dagger }
 = i   \tilde{\bar c}    X^{I \dagger } - i  X^{I \dagger } \bar c. 
\end{eqnarray}
The present analyses will be very useful in establishing the physical unitarity and
renormalizability of the ABJM theory without Higgs superfields.
{Though the  nilpotency of the
BRST symmetry leads to the physical unitarity,
  there is no general proof that the loss of nilpotency
immediately yields the violation of physical
unitarity. Therefore, even in the absence of nilpotency,
there is possibility to find another way of
proving physical unitarity. The
physical unitarity follows from the cancellation among
unphysical modes: the longitudinal and transverse modes of
the gauge field together with the Faddeev-Popov ghost and
antighost. Here we see 
  that the
violation of physical unitarity in the massive case follows from the incomplete
cancellation among unphysical modes, as, for the massive case, the physical modes are given
by a longitudinal and two transverse modes. Therefore, the remaining
unphysical mode  is not sufficient to
cancel the ghost and antighost contributions. As a result, the
elementary superfields in the original action of the ABJM model are
not sufficient to respect the physical unitarity. 
There must
be a mechanism which supplies the model with an extra
 bosonic mode.  The  non-linear superfield $B$  is propagating
in the massive case and therefore can
be an important character in the cancellation in the massive
case.}
\section{Conclusions}
 In this paper, we have studied the Faddeev-Popov quantization of the ${\cal N}=1$ ABJM theory
 in the most general covariant gauges. The absolutely anti-commuting BRST and anti-BRST
 transformations are also demonstrated.  We have constructed a massive ABJM theory
 in  ${\cal N}=1$ superspace without Higgs superfields. The presence of mass terms break  the
 BRST and anti-BRST symmetries of the theory. These  broken symmetries 
 are restored further by extending the BRST and
 anti-BRST transformations. In this context, we have found that the resulting BRST and anti-BRST symmetry transformations  lose their nilpotency due to presence of mass parameter $M$ for the superfields.
 These extended symmetries  are not absolutely anti-commuting even on the CF restricted surface and responsible candidates are the presence of non-zero mass $M$, gauge parameter $\alpha$  and   ghost-anti-ghost condensates.  Further, we have studied the
 gauge dependence of the generating functional of connected diagrams for massive ABJM model where we 
 adopt   the Kugo-Ojima subsidiary condition corresponding to extended BRST and anti-BRST transformations. Remarkably, it is observed   that the   generating functional of
the connected Green functions for massive ABJM model, $ W[J]$, depends on the parameter
$\alpha$ only if   mass   and ghost-anti-ghost condensates  are present.
Finally, we have constructed the massive gauge superfields without Higgs superfields which lead to the Proca  mass terms. The off-shell and on-shell 
extended BRST  invariant condensates are also evaluated. 
{Indeed, one can show that the norm cancellation
is automatically guaranteed from the Slavnov-Taylor identities
if the ghost-antighost bound state exists. In this way, one can recover
the physical unitarity   in a nonperturbative
way. We also would like to comment that to show the existence of the  ghost and
antighost condensate, the Nambu-Bethe-Salpeter equation is to be
solved.}
The details about the  physical unitarity and 
 renormalizability are not discussed for the  model and are the subject of future investigation.


\begin{thebibliography}{99}
\bibitem{witt}E. Witten,  Nucl. Phys. B 443, 85 (1995).
 \bibitem{abjm} O. Aharony, O. Bergman, D. L. Jafferis
 and J. Maldacena, 	JHEP 0810, 091 (2008).
 \bibitem{mari} M. Marino,   J. Phys. A: Math. Theor. 44, 463001 (2011).
 \bibitem{68}I. R. Klebanov and A. A. Tseytlin,  Nucl. Phys. B 475, 
164 (1996).
 \bibitem{32}N. Drukker, M. Marino and P. Putrov, Commun. Math. Phys. 306, 511 (2011).
 \bibitem{15} J. Bagger and N. Lambert, Phys. Rev. D 75, 0454020
(2007); 77, 065008 (2008); JHEP 02, 
 105 (2008).
\bibitem{16} A. Gustavsson, JHEP 04, 083  (2008).
\bibitem{mu} S. Mukhi and C. Papageorgakis, JHEP 05,
 085 (2008).
  \bibitem{19}   B. Ezhuthachan, S. Mukhi  and C.
Papageouragis, JHEP 04, 101 (2009).

 \bibitem{18} G. 't Hooft, Nucl. Phys. B 35, 167 (1971).
 

 \bibitem{oo} K. Hosomichi, K.-M. Lee and S. Lee, 	Phys. Rev. D 78, 066015 (2008).
 \bibitem{m} W. Nahm,   Nucl. Phys. B 135, 149 (1978).
\bibitem{n} H. Lin and J. M. Maldacena,  Phys. Rev. D 74, 084014  (2006).


 

\bibitem{kon} K. I. Kondo, Phys. Rev. D 87, 025008 (2013).
\bibitem{kon1} K. I Kondo, K. Suzuki, H. Fukamachi, S. Nishino  and T. Shinohara, Phys. Rev. D
87, 025017 (2013). 
 \bibitem{kon2} K. I Kondo,	arXiv:1202.4162 [hep-th].
 \bibitem{Mir}M. Faizal, Phys. Rev. D 84, 106011 (2011).
 \bibitem{mir1} M. Faizal and D. J. Smith, Phys. Rev. D 85, 105007  (2012). 
 \bibitem{sud} S. Upadhyay,  	Phys. Rev. D 92, 065027 (2015).
  \bibitem{sud1}M. Faizal and S. Upadhyay,  Phys. Lett. B 736, 288 (2014); S. Upadhyay, 	Int. J. Mod. Phys. A 30, 1550150 (2015).
 \bibitem{sud001} S. D. Joglekar and B. P. Mandal,   {Phys. Rev.}  { {D 51}}, 1919 (1995); S. Upadhyay, Phys. Lett. B 740, 341 (2015);  Annls.   Phys. 356, 299 (2015); Mod. Phys. Lett. A 30,1550072 (2015); Annls.  Phys. 340, 110  (2014); Annls.  Phys.  344, 290 (2014); EPL 105, 21001 (2014);  EPL  104, 61001  (2013); Phys. Lett. B 727, 293 (2013).
 \bibitem{sud2}M. Faizal, S. Upadhyay and B. P. Mandal, Phys. Lett. B 738, 201 (2014);  Int. J. Mod. Phys. A 30, 1550032 (2015);
 S. Upadhyay and D. Das, Phys. Lett. B 733, 63 (2014);
M. Faizal, B. P. Mandal and S. Upadhyay, Phys. Lett. B 721, 159 (2013). 

 \bibitem{keto} S. V. Ketov  and S. Kobayashi, {Phys. Rev.}  D 83, 045003 (2011).
  \end{thebibliography}
\end{document}